\documentclass[12pt,graphicx]{article}
\usepackage{amssymb,amsmath,amsfonts}
\usepackage{graphicx}
\usepackage{graphics}
\usepackage{eepic,epsfig}

\begin{document}
\title{Vacuum Polarization Effects on Flat Branes due to a Global Monopole}
\author{E. R. Bezerra de Mello \thanks{E-mail: emello@fisica.ufpb.br}\\
Departamento de F\'{\i}sica-CCEN\\
Universidade Federal da Para\'{\i}ba\\
58.059-970, J. Pessoa, PB\\
C. Postal 5.008\\
Brazil}
\maketitle

\begin{abstract} In this paper we analyse the vacuum polarization effects associated with a massless scalar field in the higher-dimensional spacetime. Specifically we calculate the renormalized vacuum expectation value of the square of the field, $\langle\Phi^2(x)\rangle_{Ren}$, induced by a global monopole in the "braneworld" scenario. In this context the global monopole lives in a $n=3$ dimensional sub-manifold of the higher-dimensional (bulk) spacetime, and our Universe is represented by a transverse flat $(p-1)-$ dimensional brane. In order to develop this analysis we calculate the general Green function admitting that the scalar field propagates in the bulk. Also a general curvature coupling parameter between the field and the geometry is assumed. We explicitly show that the vacuum polarization effects depend crucially on the values attributed to $p$. We also investigate the general structure of the renormalized vacuum expectation value of the energy-momentum tensor, $\langle T_{\mu\nu}(x)\rangle_{Ren.}$, for $p=3$.
\end{abstract}

\newpage
\renewcommand{\thesection}{\arabic{section}.}
\section{Introduction}
In recent years the braneworld model has received renewed interest. By this scenario our world is represented by a four-dimensional sub-manifold, a three-brane, embedded in a higher dimensional spacetime \cite{Akama,Rubakov}. The idea that our Universe may have more than four dimensions was proposed by Kaluza \cite{Kaluza} many years ago. By the Kaluza's conjecture, the gauge theories can be unified to the gravitation in a geometric formalism. Enlarging the number of dimensions of the spacetime, it is possible to accommodate the degrees of freedom associated with the gauge field in the new components of the metric tensor. In the so-called Randall-Sundrum (RS) models \cite{RS,RS1}, the spacetime contains two (RSI), respectively one (RSII), Ricci-flat brane(s) embedded on a five-dimensional Anti-de Sitter (AdS) bulk. It is assumed that all matter fields are confined on the branes and gravity only propagates in the five dimensional bulk. 

In the RSI model, the hierarchy problem between the Planck scale and the electroweak one is solved if the distance between the two branes is about $37$ times the AdS radius. Apart from the hierarchy problem, one of the most important problem in the modern physics is the cosmological constant problem (see, for instance, Ref. \cite{CP}), and many attempts addressed to this fine-tuning issue have been published. The braneworld models provide some alternative discussions about this subject. In this way, the Casimir energy associated with quantum bulk fields which obey specific boundary conditions on the branes may contribute to both, the brane and bulk cosmological constant. The Casimir energy associated with scalar field on the five-dimensional Randal-Sundrum model are calculated in \cite{NO}. Surface Casimir densities and induced cosmological constant on the branes are calculated in \cite{Sah} for a massive scalar field obeying Robin boundary conditions on two parallel brane in a general $(D+1)-$dimensional anti-de Sitter bulk.

Although topological defects have been first analysed in four-dimensional spacetime \cite{VS}, they have been considered in the context of braneworld. In this scenario the defects live in a $n-$dimensions submanifold embedded in a $(4+n)-$dimensional Universe. The domain wall case, with a single extra dimension, has been considered in \cite{Rubakov}. More recently the cosmic string case, with two additional extra dimensions, has been analysed \cite{Cohen,Ruth}. For the case with three extra dimensions, the 't Hooft-Polyakov magnetic monopole has been numerically analysed in \cite{Roessl,Cho}. In Refs. \cite{Ola} to \cite{Cho2} numerical analysis of global monopole are presented. Specifically in \cite{Cho2} it is shown that if, $\eta_0$, the energy scale where the gauge symmetry of the global system is spontaneously broken is smaller than the Planck mass, the seven-dimensional Einstein equations admit a solution which for points outside the global monopole's core is expressed by
\begin{eqnarray}
\label{g}
	ds^2=\eta_{\mu\nu}dx^\mu dx^\nu+dr^2+\alpha^2r^2d\Omega_2^2=g_{MN}d^M dx^N \ ,
\end{eqnarray}
where $\eta_{\mu\nu}=diag(-1 \  , \ 1 \ , \ ... \ , \ 1)$ is the Minkowski metric and $\alpha^2= 1-\kappa^2\eta^2_0$ a parameter smaller than unity. The solid angle deficit associated with this geometry is $\Delta\Omega=4\pi^2\kappa^2\eta^2_0$, and the critical symmetry-breaking scale is $\eta_c=1/\kappa$; moreover, this metric represents a three-dimensional global monopole with its core on the flat three brane. 

In fact the seven-dimensional action associated with this model \cite{Cho2} is:
\begin{equation}
S=\int d^7x\sqrt{-g}\left[\frac{R}{2\kappa^2}-\frac12(\partial_M \phi^a)(\partial^M\phi^a)-
\frac\lambda4(\phi^a\phi^a-\eta^2_0)^2\right]
\end{equation}
with $\kappa^2=1/M^{n+2}$, $M$ being the seven-dimensional Planck mass, and $\phi^a$ the scalar triplet associated with the global monopole. 

Adopting the most general static seven-dimensional metric tensor, with spherical symmetry  in the extra three dimensions, this tensor is given by the following line element
\begin{equation}
\label{7d}
d\hat{s}^2=B^2(r){\bar{g}}_{\mu\nu}dx^\mu dx^\nu+dr^2+C^2(r)r^2\left(d\theta^2+\sin^2\theta d\phi^2\right) =g_{AB}dx^Adx^B \ ,
\end{equation}
and for the matter field the {\it hedgehog} configuration 
\begin{equation}
\phi^a(x)=\eta_0 f(r)\hat{x}^{a+3} \ ,
\end{equation}
the solutions to the gravitational and matter field equations can be asymptotically expressed by (\ref{g}) and by $f(r)\approx 1$, respectively.  In order to be more precise, in \cite{Ola} the authors have obtained the solution to the Einstein equations considering a general $p-$dimensional Minkowski brane worldsheet and a $n\geq 3$ global monopole in the transverse extra dimensions. In this general case the metric is a generalization of (\ref{g}) with $\mu$ , $\nu= \ 0 \ , 1 \ ... \ , \ p-1$ and $\alpha^2=\frac{\kappa^2\eta^2_0}{n-2}$. 

Continuing in the analysis of quantum effects produced by fields in braneworld scenery, in this paper we shall analyse the vacuum polarization effects associated with a quantum massless scalar field propagating in a $(p+3)-$dimensional bulk spacetime which has the structure of a $p-$dimensional Minkowiski brane with a global monopole in the transverse three-dimensional sub-manifold, having the monopole's core on the brane. Our interest is to investigate the quantum effects produced by the scalar quantum  field on the brane. Although the physical interesting case is for $p=4$, we shall develop our formalism considering an arbitrary value to $p$ and make applications for specific values of it. In this paper we shall consider that the monopole is a point-like defect. The geometry associated with this spacetime is described by the metric tensor given in the line element below to the whole space:
\begin{eqnarray}
\label{g1}
	ds^2=\eta_{\mu\nu}dx^\mu dx^\nu+dr^2+\alpha^2r^2d\Omega_{(2)}^2 \ ,
\end{eqnarray}
with $\mu \ , \ \nu= \ 0 \ , 1 \ ... \ , \ p-1$ and $\alpha^2=1-\kappa^2\eta^2_0$. 
 
The main objective of this paper is to analyse the vacuum polarization effects associated with a massless scalar quantum field in the geometry given by the line element above. Specifically we shall calculate renormalized vacuum expectation value of the square of the field, $\langle\Phi^2(x)\rangle_{Ren}$, and explicitly show the behavior of this quantity for different values of $p$. We also analyse the renormalized vacuum expectation value of the energy-momentum tensor, $\langle T_{AB}(x)\rangle_{Ren.}$. In order to develop these investigations we shall calculate the Euclidean scalar Green function by making a Wick rotation $t=i\tau$ on the temporal coordinate. This function must obey the non-homogeneous second order differential equation
\begin{equation}
\label{B}
\left(\Box-\xi R\right)G_E(x,x')=-\delta^D(x,x')=-\frac{\delta^D(x-x')}{\sqrt{g}} \ ,
\end{equation}
with
\begin{eqnarray}
	\Box=\frac1{\sqrt{g}}\partial_A[{\sqrt{g}}g^{AB}\partial_B] \ .
\end{eqnarray}
We have introduced in (\ref{B}) an arbitrary curvature coupling $\xi$. Moreover, $\delta^D(x,x')$ is the bidensity Dirac distribution and $R$ is the scalar curvature. 

This paper is organized as follows: In section $2$ we construct, for an arbitrary value of $p$, the Euclidean scalar Green function for the spacetime defined by (\ref{g1}) considering an arbitrary curvature coupling $\xi$. We shall see that this Green function  is expressed in terms of an infinity sum of product of the associated Legendre functions with Legendre polynomials. In section $3$, we calculate explicitly the renormalized vacuum expectation value of the square of the scalar field, $\langle\Phi^2(x)\rangle_{Ren.}$, for specific values of $p$. In section $4$ we present the formal expression to the renormalized vacuum expectation value to the energy-momentum tensor for $p=3$ only. In section $5$ we present our conclusions and the most important remarks about this paper. In Appendix \ref{A1} we present the exact result to the renormalized vacuum expectation value of the square of the field for the case where $p=3$ and $\xi=1/8$. In Appendix \ref{A2} we also present the renormalized vacuum expectation value of the square of the field for the case where $p=4$, considering the monopole as a four-dimensional ($n=4$) topological object.

\section{The Scalar Green Function}
\label{Scalar}
In this section we calculate the Euclidean Green function associated with a massless scalar quantum field in $D=(p+3)-$dimensional spacetime defined by (\ref{g1}) admitting an arbitrary curvature coupling constant. This Green function  must obey the non-homogeneous second order differential equation 
\begin{equation}
\left(\Box-\xi R\right)G_E^{(p)}(x,x')=-\delta^D(x,x')=-\frac{\delta^p(x-x')\delta(r-r')\delta(\theta-\theta')\delta(\phi-\phi')}{\alpha^2r^2\sin\theta}  \ .
\end{equation}

In order to calculate this function we adopt the Schwinger-DeWitt formalism as shown below:
\begin{equation}
\label{Heat}
G_E^{(p)}(x,x')=\int^\infty_0 ds K(x,x';s) \ ,
\end{equation}
where the heat kernel, $K(x,x';s)$, can be expressed in terms of eigenfunctions of the operator $\Box-\xi R$ as follows:
\begin{equation}
\label{Heat-1}
K(x,x';s)=\sum_\sigma \Phi_\sigma(x)\Phi_\sigma^*(x')
\exp(-s\sigma^2) \ ,
\end{equation}
$\sigma^2$ being the corresponding positively defined eigenvalue. Writing
\begin{equation}
\left(\Box-\xi R\right)\Phi_\sigma(x)=-\sigma^2 \Phi_\sigma(x) \ ,
\end{equation}
we obtain the complete set of normalized solutions of the above equation
\begin{equation}
\Phi_\sigma(x)=\frac{{\sqrt{q}}e^{-ikx}J_{\nu_l}(qr)Y_{lm}(\theta, \ \phi)}{\alpha (2\pi)^{p/2}{\sqrt{r}}} \ ,
\end{equation}
with
\begin{equation}
\sigma^2=k^2+q^2 \ ,
\end{equation}
being $J_\nu$ the Bessel function of order
\begin{equation}
\label{nu}
\nu_l=\frac1\alpha\sqrt{\left(l+1/2\right)^2+2(1-\alpha^2)(\xi-1/8)} \ ,
\end{equation} 
So according to (\ref{Heat-1}) the heat kernel is given by the following expression:
\begin{eqnarray}
\label{Heat1}
K(x,x';s)&=&\int \ d^pk \ \int^\infty_0 dq\sum_{l,m}\Phi_\sigma(x)\Phi^*_\sigma(x')e^{-s\sigma^2} \ .
\end{eqnarray}
By using \cite{Grad}, and the addition theorem for the spherical harmonics, we obtain:
\begin{eqnarray}
\label{Heat2}
K(x,x';s)=\frac1{2^{p+3}\pi^{p/2+1}}\frac1{\alpha^2\sqrt{rr'}}\frac{e^{-\frac{\rho^2}{4s}}}{s^{p/2+1}}\sum_{l}(2l+1)I_{\nu_l} \left(\frac{rr'}{2s}\right)P_l(\cos\gamma) \ ,
\end{eqnarray}
where
\begin{eqnarray}
	\rho^2=\Delta x^2+r^2+r'^2 \ ,
\end{eqnarray}
$I_\nu$ being the modified Bessel function and $\gamma$ the angle between the two arbitrary directions on the transverse three-dimensional submanifold.

Before to go on in our calculation we would like to call attention to the fact that for $\alpha=1$, $\nu_l$ becomes $l+1/2$ and it is possible to obtain a closed expression to the sum in (\ref{Heat2}) \cite{Ab}. Consequently the closed expression to the heat kernel is
\begin{eqnarray}
	K(x,x';s)=\frac1{2^{p+3}\pi^{p/2+3/2}}\frac{e^{-\frac{\Delta X^2}{4s}}}{s^{p/2+1}} \ ,
\end{eqnarray}
where
\begin{eqnarray}
	\Delta X^2=\Delta x^2+({\vec{r}}-{\vec{r}}')^2 \ .
\end{eqnarray}

Now we are in position to obtain the Euclidean Green function by substituting (\ref{Heat2}) into (\ref{Heat}). Our final result is \cite{Pru}:
\begin{eqnarray}
\label{Green}
G^{(p)}(x,x')=\frac1{2^{\frac{p+5}2}\pi^{\frac{p+3}2}}\frac{i^{1-p}}{\alpha^2}\frac1{(rr')^{\frac{p+1}2}}\frac1{(\sinh u)^{\frac{p-1}2}}\sum_{l=0}^\infty(2l+1)Q_{\nu_l-1/2}^{\frac{p-1}2}(\cosh u)P_l(\cos\gamma) \ ,
\end{eqnarray}
where
\begin{equation}
\label{u}
\cosh u=\frac{\Delta x^2+r^2+r'^2}{2rr'} \ 
\end{equation}
and $Q^\lambda_\nu$ is the associated Legendre function. It is possible to express this function in terms of hypergeometrical functions \cite{Grad} by:
\begin{eqnarray}
Q^\lambda_\nu(\cosh u)&=&e^{i\lambda\pi}2^\lambda{\sqrt{\pi}} \ \frac{\Gamma(\nu+\lambda+1)}{\Gamma(\nu+3/2)} \frac{e^{-(\nu+\lambda+1)u}}{(1-e^{-2u})^{\lambda+1/2}}(\sinh u)^\lambda\times \nonumber\\
&&F\left(\lambda+1/2 \ , \ -\lambda+1/2 \ ; \ \nu+3/2 \ ; \ \frac1{1-e^{2u}}\right) \ .
\end{eqnarray}
For our case we must change the parameter $\nu$ by $\nu_l-1/2$ and take $\lambda=(p-1)/2$. So the relevant hypergeometric function is
\begin{eqnarray}
	F\left(\frac p2 \ , \ -\frac{p-2}2 \ ; \ \nu_l+1 \ ; \ \frac1{1-e^{2u}}\right) \nonumber \ .
\end{eqnarray}
If $p$ is a even number, this function becomes a polynomial of degree $\frac{p-2}2$; however being $p$ an odd number, this function is an infinite series.

Having obtained the Green function it is possible now to calculate the vacuum polarization effects associated with a scalar field on the bulk. We shall do this analysis in next section.

\section{The Computation of $\langle\Phi^2(x)\rangle_{Ren.}$}\label{3}
The vacuum expectation value of the square of the scalar field is formally expressed by taking the coincidence limit of the Green function as shown below:
\begin{equation}
\langle\Phi^2(x)\rangle=\lim_{x'\to x}G^{(p)}(x,x') \ .
\end{equation}
However this procedure provides a divergent result\footnote{In this case, it is a consequence of the evaluation of the associated Legendre function at unity.}. In order to obtain a finite and well defined result to this vacuum expectation value, we must apply some renormalization procedure. Here we shall adopt the point-splitting renormalization one. The basic idea of this procedure consists to analyse the divergent contributions of the Green function in the coincidence limit and subtract them off. In \cite{Wald}, Wald observed that the singular behavior of the Green function has the same structure as given by the Hadamard one, which on the other hand can be written in terms of the square of the geodesic distance between two points. So, here we shall adopt the following prescription: we subtract from the Green function the Hadamard one before applying the coincidence limit as shown below:
\begin{equation}
\label{Phi2}
\langle\Phi^2(x)\rangle_{Ren.}=\lim_{x'\to x}\left[G^{(p)}(x,x')-G_H(x,x')\right] \ .
\end{equation}
Because the explicit expression to the Hadamard function depends on the dimension of the spacetime, the above calculation can be explicitly performed by specifying values to $p$. So in the subsections that follows we shall adopt specific value to this parameter.

\subsection{Case $p=1$}
The case $p=1$ represents a very specific situation. The bulk spacetime corresponds to the four-dimensional global monopole spacetime \cite{BV}. The calculations of the vacuum polarization effects associated with massless scalar field on this spacetime has been developed by Mazzitelli and Lousto \cite{ML} long time ago. More recently the the effects of temperature on these polarization effects were analyzed in \cite{Mello1}. Moreover, the vacuum polarization effects associated with massless fermionic field on this manifold have been also analysed in \cite{Mello2}, and the effects of temperature was considered in \cite{Mello3}. The Casimir energy associated with massive scalar field inside a spherical region in the global monopole background have been analyzed in \cite{Bordag} and \cite{Mello4}, using the zeta function regularization procedure. In \cite{Mello5} it is investigated some Casimir densities associated with massive fermionic field obeying the MIT bag boundary condition on spherical shell in the global monopole spacetime. 

\subsection{Case $p=2$}\label{p2}
The case $p=2$ is a new one. It corresponds to a two-dimensional Minkowiski brane with a global monopole on the transverse sub-manifold. For this case the associated Legendre function in (\ref{Green}) assumes a very simple expression
\begin{eqnarray}
	Q_{\nu_l-1/2}^{1/2}(\cosh u)=i{\sqrt\frac\pi2}\frac{e^{-\nu_lu}}{{\sqrt{\sinh u}}} \ .
\end{eqnarray}
Consequently the Green function becomes
\begin{eqnarray}
\label{G2}
	G^{(2)}(x,x')=\frac1{16\pi^2}\frac1{\alpha^2(rr')^{3/2}}\frac1{\sinh u}\sum_l(2l+1)e^{-\nu_lu}P_l(\cos\gamma) \ .
\end{eqnarray}
Because our main objective is to calculate the vacuum expectation value of the square of the scalar field, let us take first the coincidence limit in the angular part of (\ref{G2}). Doing this the angle $\gamma$ vanishes. Unfortunately, because in general, the dependence of $\nu_l$ with $l$ in (\ref{nu}) is not simple due the presence of the parameter $\alpha\neq 1$, it is not possible to develop the summation in the quantum number $l$ and obtain a closed expression to (\ref{G2})\footnote{For the case where $\xi=1/8$ the dependence of $\nu_l$ with $l$ becomes much simpler and the summation in $l$ in (\ref{G2}) can be easily developed. This special situation will be considered separately in the Appendix \ref{A1}.} However following the Mazzitelli and Lousto procedure \cite{ML}, it is possible to obtain an approximation expression to $\nu_l$ developing a series expansion in powers of the parameter $\eta^2=1-\alpha^2$ considered smaller than unity. Up to the first order in $\eta^2$ we have:
\begin{eqnarray}
\nu_l\approx(l+1/2)(1+\eta^2/2)+\frac{(2\xi-1/4)}{2l+1}\eta^2 +O(\eta^4) \ .	
\end{eqnarray}
In this way
\begin{eqnarray}
\label{S}
S(u)=\sum_{l=0}^\infty(2l+1)e^{-\nu_lu}\cong\frac{\cosh(u/2)}{2\sinh^2(u/2)}\left[1-\frac{u\eta^2}{\sinh u}\left(1+4\xi\sinh^2(u/2)\right)\right]+O(\eta^4) \ .
\end{eqnarray}
Consequently an approximated expression to the Green function, Eq. (\ref{G2}), up to the first order in $\eta^2$, can be provided by
\begin{eqnarray}
\label{Gr2}
	G^{(2)}(x,x')=\frac1{64\pi^2}\frac1{(rr')^{3/2}}\frac{(1+\eta^2)}{\sinh^3u}\left[1-\frac{u\eta^2}{\sinh u}\left(1+ 4\xi\sinh^2(u/2)\right)\right] \ .
\end{eqnarray}

In \cite{Chr}, Christensen has given the general expression for the Hadamard function for any dimensional spacetime. There is explicitly shown that a logarithmic term in the expansion of the Hadamard function appears for a even dimensional spacetime. 

Following \cite{Chr}, we write down the Hadamard function for the massless scalar function when the dimension of the spacetime is an odd number. This functions is given by
\begin{equation}
G_H(x,x')=\frac{\Delta^{1/2}(x,x')}{2(2\pi)^{n/2}}\frac1{\sigma^{n/2-1}(x,x')}\sum_{k=0}^{\frac{n-3}2}a_k(x,x')\sigma^k(x,x')\frac{\Gamma(n/2-k-1)}{2^k} \ ,
\end{equation}
where $2\sigma(x,x')$ is the square of the geodesic distance. $\Delta(x,x')$, the Van Vleck-Morette determinant and the coefficients, $a_k(x,x')$, for $k=0, \ 1,\ 2$, have been computed by many authors. (See Refs. \cite{BS} and \cite{Chr1}.)

In this case $n=5$, so the Hadamard function reads:
\begin{eqnarray}
\label{H1}	G_H(x,x')=\frac{\Delta^{1/2}(x,x')}{2^{9/2}\pi^2}\frac1{\sigma^{1/2}(x',x)}\left[\frac1{\sigma(x',x)}+\left(\frac16-\xi\right)R(x)\right] \ ,
\end{eqnarray}
being $R(x)$ the curvature scalar which for this spacetime, up to the first order in $\eta^2$, is given by
\begin{eqnarray}
	R(x)=\frac{2\eta^2}{r^2} \ .
\end{eqnarray}
In this approximation we also can use $\Delta^{1/2}(x,x')=1$.

Using (\ref{u}), it is possible to express the one-half of square of geodesic distance $\sigma(x,x')$ in (\ref{H1}) by $2(rr')\sinh^2(u/2)$. Now substituting (\ref{Gr2}) and (\ref{H1}) into (\ref{Phi2}), we obtain, after a long calculation, a vanishing result to the vacuum polarization effect, i.e., we get
\begin{eqnarray}
\label{Po}
\langle\Phi^2(x)\rangle_{Ren.}=0 \ .	
\end{eqnarray}

In previous paper \cite{Mello6}, we have calculated the renormalized vacuum expectation value of the square of a scalar field in higher dimensional global monopole spacetime, i.e., a spacetime of $(1+d)-$dimensions, with $d\geq3$, in which the global monopole lives in whole space. For this spacetime we have obtained that for the case $d=4$,  $\langle\Phi^2(x)\rangle_{Ren.}$ presents a non-vanishing result. It fact it is proportional to the inverse of the third power of the distance from the point to the monopole's core.

Because the above vanishing result, we present below the calculation of the vacuum polarization effect, $\langle\Phi(x)\rangle_{Ren.}$, considering the next to leading order contribution, i.e., the contribution proportional to $O(\eta^4)$. To do that we have to construct the Green function in this order. The obtained result can be expressed by:
\begin{eqnarray}
{\bar{G}}(x,x')=\frac1{16\pi^2}\frac1{(rr')^{3/2}}\frac1{\sinh u}[S_0(u)+S_1(u)+S_2(u)]\eta^4+O(\eta^6) \ ,	
\end{eqnarray}
where $S_0(u)$ and $S_1(u)$ correspond to the terms independent of $\eta$ and that multiplies $\eta^2$ in (\ref{S}), respectively. The term $S_2(u)$, is the new one. It is obtained by expanding $\nu_l$ up to $\eta^4$ order and develop the geometric summation. It reads
\begin{eqnarray}	S_2(u)&=&\left\{\frac{u^2}2\left[\frac{(\cosh(3u/2)+23\cosh(u/2))}{128\sinh^4(u/2)}+\frac{(\xi-1/8)}2\frac{\cosh(u/2)}{\sinh^2(u/2)}\right.\right.
\nonumber\\
&-&\left.\frac{(2\xi-1/4)^2}2\ln\left(\tanh(u/2)\right)\right]-u\left[\frac3{64}\frac{\cosh(u)}{\sinh^3(u/2)}+ \frac9{64}\frac1{\sinh^3(u/2)}\right. \nonumber\\
&+&\left.\left.\frac{(\xi-1/8)}2\frac1{\sinh(u/2)}-(\xi-1/8)^2f(u)\right]\right\}\eta^4 \ ,
\end{eqnarray}
where $f(u)$ is a long expression given by the integrals of $\ln(\tanh(u/2))$. In order to calculate the renormalized vacuum expectation value of the square of the field, we have also to construct the Hadamard function at order $\eta^4$ too. These contributions come from the scalar curvature, $R(x)$, and $a_1(x)$ coefficient. Finally substituting the results obtained into (\ref{Phi2}) we get a non-vanishing result:
\begin{eqnarray}
\langle\Phi(x)\rangle_{Ren.}=-\frac{\eta^4}{192r^3}\left(\xi-\frac18\right) \ .	
\end{eqnarray}

\subsection{Case $p=3$}
\label{p3}
For $p=3$ the associated Legendre function in (\ref{Green}) is an infinite series. So, in order to investigate the vacuum polarization effect we shall use the integral representation given below \cite{Grad}:
\begin{eqnarray}
\label{Q1}
Q_{\nu-1/2}^\lambda(\cosh u)={\sqrt{\frac\pi 2}}\frac{e^{i\lambda\pi}\sinh^\lambda(u)}{\Gamma(1/2-\lambda)}\int_u^\infty \ dt \ \frac{e^{-\nu t}}{(\cosh t- \cosh u)^{\lambda+1/2}} \ .
\end{eqnarray}
For this case $\lambda=(p-1)/2=1$. However the above representation can only be applied for $Re(\lambda)<1/2$.  This integral representation, on the other hand, can be used for submanifold $(p-1)-$brane of smaller dimension. In the calculation of vacuum polarization effects, we have adopted the point-splitting renormalization procedure, subtracting from the Green function the Hadamard one. This procedure provides a finite and well defined result to evaluate the renormalized vacuum expectation value of the square of the scalar field. In what follows, we shall allow in this renormalization procedure, that the dimension of the brane be an arbitrary number. In this way we may use (\ref{Q1}) in Green function (\ref{Green}), and also in the definition of Hadamard function. Finally, in the calculation of the vacuum polarization effect, we shall take $p\to 3$ before to take the coincidence limit in the renormalized Green function. As we shall see we shall obtain a finite and well defined result\footnote{A similar procedure has been used in sub-section $(6.2)$ of \cite{BD} to compute the singular geometrical part of the effective action associated with scalar field in curved space.}. Adopting this procedure the Green function can be written by
\begin{eqnarray}
\label{Gr3}
	G^{(3)}(x',x)&=&\frac{{\sqrt{2}}}{32\pi^2{\sqrt{\pi}}}\frac1{\alpha^2(rr')^2}\frac1{\Gamma(1/2-\lambda)}\int_u^\infty\frac{dt} {(\cosh t-\cosh u)^{\lambda+1/2}}\times\nonumber\\
&&\sum_{l=0}^\infty(2l+1)e^{-\nu_lt}P_l(\cos\gamma) \ .
\end{eqnarray}
Taking $\gamma=0$ into the above equation it is possible to develop an approximated expression to the summation on the angular quantum number $l$, as we did in (\ref{S}).

The Hadamard function in a six dimensional spacetime reads:
\begin{eqnarray}
\label{H2}	G_H(x',x)=\frac{\Delta^{1/2}(x,x')}{16\pi^3}\left[\frac{a_0(x,x')}{\sigma^2(x,x')}+\frac{a_1(x,x')}{2\sigma(x,x')}-\frac{a_2(x,x')}4\ln\left(\frac{\mu^2\sigma(x,x')}2\right)\right] \ ,
\end{eqnarray}
where $\mu$ is an arbitrary energy scale. For the radial point-splitting, $\Delta x=0$, we have $\sigma(x',x)= (r'-r)^2/2$. Now substituting the expressions to the coefficient $a_k$, for $k=0, \ 1,\ 2$, we get, up to the first order expansion in the parameter $\eta^2$, the following expression:
\begin{eqnarray}
\label{H22}
G_H(r',r)=\frac1{16\pi^3}\left[\frac4{(r'-r)^4}+\frac{2\eta^2}{r^2}\left(\frac16-\xi\right)\frac1{(r'-r)^2}
-\frac{\eta^2}{6r^4}\left(\frac15-\xi\right)\ln\left(\frac{\mu^2(r'-r)^2}4\right)\right]	\ .
\end{eqnarray}
At this point we shall adopt the same approach used to construct the Green function. We shall express the different powers of $\frac1{r'-r}$ in the Hadamard function above by the following integral representation:
\begin{eqnarray}
\label{Int}
	\frac1{(r'-r)^{d+1}}&=&\frac{(r'-r)^{2(\lambda-1)}}{2^{d+\lambda-\frac32}}\frac1{(r'r)^{\frac{d+2\lambda-1}2}}\frac{\Gamma(\frac d2)}{\Gamma(\frac{d-1}2+\lambda)\Gamma(\frac12-\lambda)}\times\nonumber\\
&&	\int_u^\infty\frac{dt}{(\cosh t-\cosh u)^{\frac12+\lambda}} \frac{\cosh(t/2)}{\sinh^{d-1}(t/2)} \ ,
\end{eqnarray}
for $d\geq1$. Substituting the parameter $d$ for the appropriated values in order to reproduce the correct powers of $\frac1{r'-r}$ in (\ref{H22}), and expressing the logarithmic term by $Q_0(\cosh u)$ we obtain a long expression. Now we have to substitute (\ref{Gr3}) and (\ref{H22}) into (\ref{Phi2}), in order to obtain the renormalized vacuum expectation value to the square of the field. However, as we have mentioned before, we shall take $\lambda\to1$ first into the renormalized Green function before to take the coincidence limit. Doing this procedure we get\footnote{We have used the program MAPLE version $9.5$ to perform the integrals.}:
\begin{equation}
\langle\Phi^2(x)\rangle_{Ren.}=\frac1{576\pi^3}\frac{\eta^2}{r^4}\left(\frac{47}{25}-10\xi\right)+\frac1{48\pi^3}\frac{\eta^2}{r^4}\left(\xi-\frac15\right)\ln(\mu r) \ .
\end{equation}
We can see that for the conformal coupling in six dimension, $\xi=1/5$, there is no ambiguity in the definition of the above vacuum polarization effect, i.e., the logarithmic contribution disappears and we get $\langle\Phi^2(x)\rangle_{Ren.}=\frac{11}{7200\pi^3}\frac{\eta^2}{r^4}$. Now at this point we may compare the above result with the similar one obtained in \cite{Mello6} for a six-dimensional global monopole spacetime. We can observe that, although the calculations have been developed in different spacetimes, the results present some similarity\footnote{In \cite{Mello6} we have found $\langle\Phi^2(x)\rangle_{Ren.}=-\frac{\eta^2}{96\pi^3r^4} \left(\frac{47}{25}-10\xi\right)+\frac{\eta^2}{8\pi^3r^4}\left(\xi-1/5\right)\ln(\mu r)$.}: By dimensional arguments we expected that they present contributions proportional to the inverse of the fourth power of $r$, and because both spacetime have a non-vanishing scalar curvature, they also present logarithmic contributions that disappear for $\xi=1/5$. The new fact that seems interesting to us is that, up to the first order in the parameter $\eta^2$, the term inside the parenthesis are the same in both calculations, i.e., for $\xi=47/250$, the contributions proportional to $1/r^4$ disappear\footnote{In the Appendix \ref{A1} we shall see that for $\xi=1/8$ it is possible to provide an exact result to $\langle\Phi^2(x)\rangle_{Ren.}$.}.
 
\subsection{Case $p=4$}
In this subsection we shall analyse the vacuum polarization effects due to the global monopole considering that the brane is a flat four-dimensional submanifold. Although in the other subsections we have analysed the quantum effect from theoretical point of view, here the analysis seems more relevant to our world. For $p=4$ the Legendre function in (\ref{Green}) can be expressed by a simple expression
\begin{eqnarray}
	Q^{3/2}_{\nu_l-1/2}(\cosh u)=-i\sqrt{\frac\pi{2\sinh u}}\ (\nu_l+1)e^{-\nu_lu}\left[1+\frac{e^{-u}} {(\nu_l+1)\sinh u}\right] \ .
\end{eqnarray}
Taking $\gamma=0$ and substituting the above expression into (\ref{Green}), we get:
\begin{eqnarray}
	G(x',x)=\frac1{32\pi^3}\frac1{\alpha^2(r'r)^{5/2}}\frac1{\sinh^2u}\left[\sum_{l=0}^\infty (2l+1)(\nu_l+1)e^{-\nu_l u}+\frac{e^{-u}}{\sinh u}\sum_{l=0}^\infty (2l+1)e^{-\nu_lu}\right] \ .
\end{eqnarray}

Here again it is possible to provide an approximated expression to the Green function above by developing an expansion in powers of the parameter $\eta^2$. The term inside the bracket can be written by $S(u)\coth u-S'(u)$, where $S(u)$ is given by (\ref{S}) and the prime denotes differentiation with respect to $u$. Developing the expansion up to the first order in $\eta^2$ we have:
\begin{eqnarray}
\label{Gr4}
	G^{(4)}(x',x)=\frac3{512\pi^3}\frac{1+\eta^2}{(r'r)^{5/2}}\frac1{\sinh^5(u/2)}+\frac1{32\pi^3}\frac1{(r'r)^{5/2}}\frac1{\sinh^2u} \left[ {\cal{V}}_1(u)+\xi{\cal{V}}_2(u)\right]\eta^2 \ ,
\end{eqnarray}
where ${\cal{V}}_1(u)$  and ${\cal{V}}_2(u)$ are two long expressions. 

The Hadamard function for a seven-dimensional spacetime reads:
\begin{eqnarray}
	G_H(x',x)=\frac{\Delta^{1/2}(x',x)\sqrt{2}}{128\pi^3}\left[\frac{3a_0(x',x)}{\sigma^{5/2}(x',x)}+\frac{a_1(x',x)}{\sigma^{3/2}(x',x)}+ \frac{a_2(x',x)}{\sigma^{1/2}(x',x)}\right] \ .
\end{eqnarray}
Expressing the geodesic distance $\sigma(x',x)$ by $2rr'\sinh^2(u/2)$ and developing, up to the first order in $\eta^2$, the coefficients $a_k(x',x)$ for $k= \ 1, \ 2, \ 3 $ and $\Delta^{1/2}(x',x)$, we have:
\begin{eqnarray}
\label{H3}
	G_H(x',x)&=&\frac3{512\pi^3}\frac{1}{(r'r)^{5/2}}\frac1{\sinh^5(u/2)}+\frac1{128\pi^3}\frac{\eta^2}{r^2}\left(\frac16 -\xi\right)\frac1{(r'r)^{3/2}}\frac1{\sinh^3(u/2)}\nonumber\\
	&+&\frac1{192\pi^3}\frac{\eta^2}{r^4}\left(\frac15 -\xi\right) \frac1{(r'r)^{1/2}}\frac1{\sinh(u/2)} \ .
\end{eqnarray}

Substituting (\ref{Gr4}) and (\ref{H3}) into (\ref{Phi2}), and after some intermediate steps we obtain
\begin{eqnarray}
\langle\Phi^2(x)\rangle_{Ren.}=0 \ .	
\end{eqnarray}

This result, together with (\ref{Po}), explicitly shown that the renormalized vacuum expectation values of the square of the scalar field induced by a global monopole is zero up to the first order in $\eta^2$ when $p$ is equal to $2$ and $4$. For these cases we can observe that the respective Green functions are odd functions of $u$. So, taking the limit $u\to 0$ no terms independent of $u$ are left. Applying the renormalization procedure, all singular terms are canceled and the regular ones go to zero. In subsection (\ref{p2}), we have shown that considering next to leading order term a non-vanishing result for $\langle\Phi(x)\rangle_{Ren.}$ has been obtained. Adopting a similar procedure in this subsection, i.e., constructing the Green and Hadamard functions for this seven dimensional spacetime at order $\eta^4$, we have found that this vacuum expectation value is still zero. Although we cannot affirm that this quantity remains zero for the next order of approximation, we may want to know if being the global monopole a higher dimensional topological object, the renormalized vacuum expectation value of the square of the field  becomes different from zero. In order to clarify this point, in Appendix \ref{A2} we calculate $\langle\Phi^2(x)\rangle_{Ren.}$  considering that the global monopole lives in the transverse extra four-dimensional space.

\section{The Energy-Momentum Tensor}
In this paper we are analysing the quantum effects associated with a massless scalar field in the metric spacetime defined by (\ref{g1}), which corresponds to a bulk spacetime constituted by a $(p-1)-$dimensional flat brane and a three-dimensional global monopole submanifold transverse to it. Because the metric tensor does not present any dimensional parameter, and also because we are working with natural units system ($\hbar=c=1$), we can infer that the vacuum polarization effects must depend on the radial coordinate $r$ and, due to the non-vanishing scalar curvature of the spacetime, on the renormalization mass scale, $\mu$, too. Moreover, by dimensional analysis we expect that $\langle\Phi^2(x)\rangle_{Ren.}$ and $\langle T_{AB}(x)\rangle_{Ren.}$ are proportional to $1/r^{p+1}$ and to $1/r^{p+3}$, respectively. The factors of proportionality depend on the parameter $\eta^2$ and the non-minimal coupling constant $\xi$. By the calculations developed in section \ref{3}, we have shown that the renormalized vacuum expectation value of the square of the scalar field is zero up to the first order in $\eta^2$ for two- and four-dimensional flat branes. Although we cannot affirm that these vanishing results also occur in the calculation of the renormalized vacuum expectation values of the energy-momentum tensor, here we shall analyse $\langle T_{AB}(x)\rangle_{Ren.}$ for $p=3$ only. In this case the dimension of the bulk spacetime is six. 

The renormalized vacuum expectation value of the energy-momentum tensor should obey the conservation condition, i.e.,
\begin{equation}
\label{Con}
\nabla_A\langle T_B^A(x)\rangle_{Ren.}=0 \ ,
\end{equation}
and provides the correct trace anomaly for this six-dimensional spacetime \cite{Chr}:
\begin{equation}
\label{Ano}
\langle T_A^A(x)\rangle_{Ren.}=\frac1{64\pi^3}a_3(x) \ .
\end{equation}
 
Taking into account all the above informations, we can conclude that the general structure for the renormalized vacuum expectation value of the energy-momentum tensor is:
\begin{equation}
\label{T}
\langle T_B^A(x)\rangle_{Ren.}=\frac1{64\pi^3r^6}\left[F_B^A(\eta^2\xi)+G_B^A(\eta^2,\xi)\ln(\mu r)\right] \ .
\end{equation}
The components of the tensor $F_B^A$ obey specific restriction that will be examined later. As to the tensor $G_B^A$, it is possible to provide an expression to it, up to the first powers in the parameter $\eta^2$.

Because the presence of the arbitrary cutoff scale $\mu$, there is an ambiguity in the definition of (\ref{T}). Moreover, the change in this quantity under the change of the renormalization scale is given in terms of the tensor $G_B^A$ as shown below:
\begin{equation}
\langle T_B^A(x)\rangle_{Ren.}(\mu)-\langle T_B^A(x)\rangle_{Ren.}(\mu')= \frac1{64\pi^3r^6}G_B^A(\eta^2,\xi)\ln(\mu/\mu') \ .
\end{equation}
In Ref. \cite{Chr}, Christensen has pointed out that the difference between these expectation values is given in terms of the effective action which depends on the logarithmic terms, whose final expression in arbitrary even dimension is
\begin{equation}
\langle T_{AB}(x)\rangle_{Ren.}(\mu)-\langle T_{AB}(x)\rangle_{Ren.}(\mu')= \frac1{(4\pi)^{n/2}}\frac1{\sqrt{g}}\frac\delta{\delta g^{AB}}\int d^nx\sqrt{g}a_{n/2}(x)\ln(\mu/\mu') \ .
\end{equation}  

In this six-dimensional spacetime we need the coefficient $a_3(x)$. In the papers by Gilkey \cite{Gilkey}, and Jack and Parker \cite{JP}, the explicit expression for this coefficient can be found for a scalar second order differential operator $D^2+X$, $D_M$ being the covariant derivative including gauge field and $X$ an arbitrary scalar function. This expression involves $43$ terms. We shall not repeat it here in a complete form. The reason is because our 
calculations have been developed up to the first order in the parameter $\eta^2$, and only the quadratic terms in Riemann and Ricci tensors, and in the scalar curvature are relevant for us\footnote{The terms proportional to $\Box^2R$ in $a_3(x)$ do not contribute because they can be written as a total derivative.}. This reduces to $12$ the number of terms which will be considered. Discarding the gauge fields and taking $X=\xi R$ we get:
\begin{eqnarray}
\overline{a}_3(x)&=&\frac16\left(\frac16-\xi\right)\left(\frac15-\xi\right)
R\Box R+\frac{\xi^2}{12}R^{;M} R_{;M}+\frac\xi{90}R^{MN}
R_{;MN}-\frac{\xi}{36}R^{;M} R_{;M}
\nonumber\\
&&-\frac1{7!}\left[28R\Box R +17R_{;M} R^{;M}-2R_{MN;P}
R^{MN;P}-4R_{MN;P}R^{MP;N}\right.+
\nonumber\\
&&9R_{MNPS;G}
R^{MNPS;G}-8R_{MN}\Box R^{MN}+24R_{MN}
R^{MP;N}\ _P+
\nonumber\\
&&\left.12R_{MNPS}\Box R^{MNPS}
\right]+O(R^3) \ .
\end{eqnarray} 
This expression is of sixth order derivative on the metric tensor. Our next step is to take the functional derivative of $\overline{a_3}(x)$ with respect to the metric tensor. Using the expressions for the functional derivative of the Riemann and Ricci tensor, together with the scalar curvature \cite{BS}, we obtain after a long calculation the following expression for the tensor\footnote{In \cite{BS} the convention adopted for the Riemann tensor is different from the used here 
which makes its sign opposite to ours.} $G_B^A$:
\begin{eqnarray}
\label{G}
G_B^A(\eta^2,\xi)&=&\frac{r^6}6\left[-\delta_B^A\Box^2 R\left(\xi^2-\frac\xi3+\frac{23}{840}\right)+\frac1{140}\Box^2 R_B^A+\right.
\nonumber\\
&&\left.\nabla^A\nabla_B\Box R\left(\xi^2-\frac\xi3+\frac1{42}\right)
\right]+O(R^2) \ .
\end{eqnarray}
In order to obtain a numerical expression to (\ref{G}), we have to substitute the Ricci tensor and scalar curvature defined in this geometry. Developing all the derivatives, we obtain, after some intermediate calculations, the expression to $G_B^A(\eta^2,\xi)$. This expression, up to the first order in the parameter $\eta^2$,  is:
\begin{eqnarray}
G_A^B(\eta^2,\xi)&=-&\frac{\eta^2}{175}\ diag\ (\ 1, 1, 1, 1, -2,-2\ )-
\nonumber\\
&&8\eta^2(\xi-1/5)(\xi-2/15)\ diag\ (\ 1, 1, 1, -2/3, 4/3, 4/3\ ) \ .
\end{eqnarray}

After this analysis, let us obtain some restrictions on the components of the tensor $F^A_B$. However, before to embark in this calculation, it is crucial to have some relations between the components of the renormalized vacuum expectation value of the energy-momentum tensor. The geometric structure of the brane section of this spacetime is Minkowiski-type. For this reason the Green functions calculated depend on the variables on the brane by $\Delta x^2=\eta_{\mu\nu}\Delta x^\mu\Delta x^\nu$. On the other hand it is possible to provide an approximate expression to the geodesic distance between two arbitrary points in this six-dimensional global monopole spacetime. By using Eq. $(A6)$ of \cite{Popov} we can see that $\sigma(x',x)$ depends on the coordinate on the brane by $\Delta x^2$ too. Finally, because the renormalized vacuum expectation value of the energy-momentum tensor can be evaluated by applying the a bi-vector differential operator \cite{BD}, ${\cal{D}}_{AB'}(x,x')$, on the renormalized Green function $G(x,x')-G_H(x,x')$, we expect that $\langle T^0_0\rangle_{Ren.}=\langle T^1_1\rangle_{Ren.}=\langle T^2_2\rangle_{Ren.}$. In fact by the expression obtained to $G_B^A(\eta^2,\xi)$, we can see that $G_0^0=G_1^1=G_2^2$ for an arbitrary curvature coupling $\xi$. So admitting the equality between the components of energy-momentum tensor, we have $F_0^0=F_1^1=F_2^2$. Now we are in position to provide some relations between the other components of $F^A_B$. Because the symmetry of the system, we can infer that this tensor should be diagonal; moreover, by using the conditions (\ref{Con}) and (\ref{Ano}), it is possible to express all the components of this tensor in terms of just one component. After some intermediate steps we find
\begin{eqnarray}
	F^3_3&=&F^0_0+\frac{G_3^3}3+(G_0^0-G_3^3)\ln(\mu r)-T \ ,\\
	F_4^4&=&F^5_5=-2F_0^0-\frac{G_3^3}6+\frac{2T}3-(G_4^4+2G_0^0)\ln(\mu r) \ ,
\end{eqnarray}
being $T=\ 64\pi^3r^6\langle T_A^A(x)\rangle_{Ren.}$. For $\xi=1/5$, the above equations become
\begin{eqnarray}
	F^3_3&=&F^0_0+\frac{G_3^3}3-T \ ,\\
	F_4^4&=&F^5_5=-2F_0^0-\frac{G_3^3}6+\frac{2T}3 \ ,
\end{eqnarray}
with
\begin{eqnarray}
	T=r^6a_3(x)=r^6\frac{18}{7!}\Box^2R+O(\eta^4)=\frac{6\eta^2}{35}+O(\eta^4) \ .
\end{eqnarray}

The complete evaluation of $\langle T^A_B(x)\rangle_{Ren.}$ requires the knowledge of at least one component of $F_B^A$, say $F_0^0$; however we shall not attempt to do this straightforward and long calculation here.

\section{Concluding Remarks}
In the context of brane world scenario, we have analysed the quantum effects associated with a massless scalar field with arbitrary curvature coupling on a $(p+3)-$dimensional bulk spacetime, due to the presence of a global monopole which lives in a three dimensional submanifold. 
 
In order to develop this analysis an important quantity evaluated was the Green function. So, in this way we have constructed the bulk scalar Green function for the general case considering the dimension of the flat brane as been an arbitrary number $p$. The respective Green function, $G^{(p)}(x',x)$, is expressed in terms of an infinity sum of product of associated Legendre functions with Legendre polynomials. The case $p=1$ reduces the geometry associated with (\ref{g1}) to a  four-dimensional global monopole spacetime. Because the calculation of the vacuum polarization effect associated with a massless scalar field on this manifold has been developed by Mazzitelli and Lousto in Ref. \cite{ML}, we decided do not consider this case, and extend the calculation of vacuum polarizations to the other cases with $p=2, \ 3, \ 4$. For the $p=2$ case, the bulk spacetime has odd dimension. In this case we find, by explicit calculation, that $\langle\Phi^2(x)\rangle_{Ren.}$ acquires a non-vanishing result only at the order $\eta^4$, being $\eta$, the parameter of expansion, defined by $\eta^2=1-\alpha^2$, considered smaller than unity. The case $p=3$ provides more interesting result. The dimension of the bulk spacetime is even. This indicates that for non-conformal curvature coupling, $\xi$, there is an ambiguity in the calculation of $\langle\Phi^2(x)\rangle_{Ren.}$. In order to calculate this quantity, we have used an integral representation to the associated Legendre function $Q_{\nu_l-1/2}^{\frac{p-1}2}(\cosh u)$, considering $p$ as an arbitrary variable which can be analytically continued throughout the complex plane. The renormalized vacuum expectation value of the square of the field operator can be evaluated by removing the singular contributions which appear in the coincidence limit of the Green function. This renormalization procedure is implemented by a systematic way: we take the difference between the Green function with the Hadamard one. The later also defined in terms of integral representation which present the dimension of the brane $p$ as an arbitrary variable. The prescription adopted to calculate the vacuum polarization effect is to take first the the limit $p\to 3$, and then the coincidence limit. Doing this we get a finite result. Comparing this result with similar one obtained in a six-dimensional global monopole spacetime, we pointed out some similarities between them about their general structure. The last case analysed in the main part of this paper is for $p=4$. In this case the associated Legendre function becomes a polynomial and the Green function can be expressed in terms a geometric sum. The dimension of the bulk spacetime is also odd, and in this case we have obtained that $\langle\Phi^2(x)\rangle_{Ren.}=0$, up to the order $O(\eta^4)$. Because this vanishing result, we decide to calculate $\langle\Phi^2(x)\rangle_{Ren.}$ considering the global  monopole as been a higher dimensional topological object. So in Appendix \ref{A2} we explicitly show that considering the global monopole as a four dimensional defect, the renormalized vacuum expectation value of the square of the field gets a non-vanishing result. Moreover, although the dimension of this bulk spacetime is even, our result, developed up to the first order in the parameter $\eta^2$, does not present ambiguity, i.e., does not present term proportional to the logarithmic of an arbitrary scale parameter $\mu$.

By dimensional arguments, we have investigated the renormalized vacuum expectation value of the energy-momentum tensor, $\langle T_{AB}(x)\rangle_{Ren.}$ for the case $p=3$. We have shown that this quantity behaves as $1/r^6$, where $r$ is the distance from the monopole's core. Because the bulk spacetime presents even dimensions, there appear in this calculation an additional term proportional to $\ln(\mu r)/r^6$, being $\mu$ is an arbitrary energy scale induced by the renormalization prescription. This term is associated with the coefficient $a_3(x)$, which, according to \cite{BD} on pp. $159$, comes from the purely geometric (divergent) Lagrangian that should renormalize the gravitational Lagrangian. When this extra term is inserted into the gravitational action, the left-hand side of the field equation is modified by the presence of order six terms proportional to:
\begin{equation}
c_1 g_{AB}\Box^2 R+c_2 \Box^2 R_{AB}+c_3 \nabla_A\nabla_B \Box R +
O(R^2) \ .
\end{equation}

Finally we want to say that the model analysed here presents  monopole's core residing on the flat brane, so the quantum influence due to the scalar field on the brane can be evaluated in the region near the monopole's core. Because we have considered the monopole as a point-like object, all the quantum effects due to matter field on the brane present singularities at the monopole's core. The problem of singularity may be prevented considering a more realistic model to this object, i.e., considering a inner structure to the monopole's core. In this way the Euclidean Green function in the regions inside and outside the monopole must be obtained. It is our intention to continue this analysis in a near future.

\section*{Acknowledgment} 
The author thanks to Aram Saharian for pertinent comments, Conselho Nacional de Desenvolvimento Cient\'\i fico e Tecnol\'ogico (CNPq.) and FAPESQ-PB/CNPq. (PRONEX) for partial financial support. 

\appendix

\section{Exact Calculation of $\langle\Phi^2(x)\rangle_{Ren.}$}
\label{A1}
In this appendix we calculate the exact expression to $\langle\Phi(x)\rangle_{Ren.}$ considering a flat $2-$brane, i.e. $p=3$, and that the global monopole is a three dimensional topological object. 

For $\xi=1/8$ we have a great simplification in our calculations. For this value
\begin{eqnarray}
	\nu_l=\frac{2l+1}{2\alpha} \ .
\end{eqnarray}
In this case the sum (\ref{S}) can be developed exactly:
\begin{eqnarray}
S(t)=\sum_{l=0}^\infty(2l+1)e^{-\nu_lt}=\frac{\cosh(t/2\alpha)}{2\sinh^2(t/2\alpha)} \ .	
\end{eqnarray}
The equation (\ref{Gr3}), for $\gamma=0$, reads now
\begin{eqnarray}
\label{Gre3}	G^{(3)}(x',x)&=&\frac{{\sqrt{2}}}{64\pi^2{\sqrt{\pi}}}\frac1{\alpha^2(rr')^2}\frac1{\Gamma(1/2-\lambda)}\int_u^\infty\frac{dt} {(\cosh t-\cosh u)^{\lambda+1/2}}\times\nonumber\\
&&\frac{\cosh(t/2\alpha)}{\sinh^2(t/2\alpha)} \ .	
\end{eqnarray}

The Hadamard function in a six dimensional spacetime was given before in Eq. (\ref{H2}); however in this exact calculation we have to use, for the coefficients $a_i$, for $i=2$ and $3$, the complete expression given below
\begin{eqnarray}
	a_1(x,x')&=&\left(\frac16-\xi\right)R=\frac1{12}\frac{(1-\alpha^2)}{\alpha^2r^2} \ ,  \nonumber\\ 
	a_2(x',x)&=&-\frac1{180}(R_{AB}R^{AB}-R_{ABCD}R^{ABCD})+\frac16\left(\frac15-\xi\right)\Box R+\frac12\left(\frac16-\xi\right)^2R^2\nonumber\\
	&=&\frac{(1-\alpha^2)(17\alpha^2+7)}{480\alpha^4r^4} \ .
\end{eqnarray}
Besides we have used $a_1=\Delta=1$. 

For the radial point-splitting separation, the Hadamard function will be expressed in terms of powers of $\frac1{r'-r}$ and logarithmic of $r'-r$. Here we shall adopt the same procedure adopted in \ref{p3}. We shall use the integral representation (\ref{Int}) and the Legendre function $Q_0(\cosh u)$, to express the terms that appear in the Hadamard function. The result is a long expression. Now substituting (\ref{Gre3}) and the obtained Hadamard function into (\ref{Phi2}) we get a finite result given below:
\begin{eqnarray}
\label{PPI}
	\langle\Phi^2(x)\rangle_{Ren.}&=&-\frac1{256\pi^3}\frac1{\alpha^2r^4}\int_0^\infty\frac{dt}{\sinh^3(t/2)}\left[ \frac{\cosh(t/2\alpha)}{\sinh^2(t/2\alpha)}-\frac{\alpha^2\cosh(t/2)}{\sinh^2(t/2)}\right.\nonumber\\
	&-&\left.\frac{(1-\alpha^2)}6\cosh(t/2)+(1-\alpha^2)\frac{(17\alpha^2+7)}{120\alpha^2}\sinh^2(t/2)e^{-t/2}\right] \nonumber\\
	&+&(1-\alpha^2)\frac{(17\alpha^2+7)}{15360\pi^3\alpha^4r^4}\ln(\mu r) \ .
\end{eqnarray}
Unfortunately it is not possible to provide an analitycal result to the integral above for a general value of $\alpha$; however we can check that the integrand behaves as $\frac{(-1+\alpha^2)(17\alpha^2+7)}{120\alpha^2}+O(t)$ in the region $t\approx 0$ and goes to zero as $t\to\infty$. The dependence of the first contribution of the vacuum expectation above on the parameter $\alpha$ can only be provided numerically. Our numerical results to this quantity for a fixed $r$ are exhibted in figure $1$. \\
\begin{figure}[tbph]
\begin{center}
\epsfig{figure=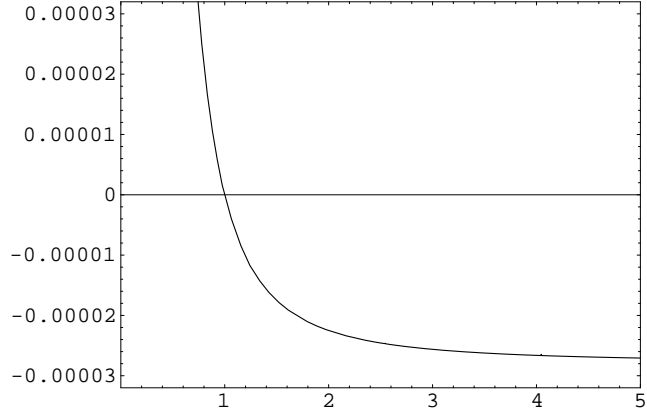,width=8.5cm,height=5.5cm} 
\end{center}
\caption{This graphs represents the first contribution of renormalized vacuum expectation value of the square of the field, (\ref{PPI}), multiplied by $r^4$ as function of the parameter $\alpha$. This vacuum expectaion value goes to infinity for $\alpha<<1$, large solid angle deficit, and  tends to a negative constant, $-0.0000278317$, for $\alpha>>1$, large solid angle excess.}
\label{fig1}
\end{figure}

\section{Case with Four-Dimensional Global Monopole}
\label{A2}
Here we consider the case where the global monopole is a four-dimensional topological object living in a  submanifold transverse to the three-dimensional flat brane. The metric tensor associated with the bulk spacetime is:
\begin{eqnarray}
	ds^2=\eta_{\mu\nu}dx^\mu dx^\nu+dr^2+\alpha^2r^2d\Omega^2_3=g_{AB}dx^Adx^B \ ,
\end{eqnarray}
where $\mu,\  \nu=0, \ 1, \ 2, \ 3$ and $x^A=(x^\mu, r, \theta_1, \theta_2, \phi)$. The coordinates are defined in the intervals as follows: $x^\mu\in(-\infty, \infty)$, $\theta_i\in [0, \pi]$ for $i=1, \ 2$ and $\phi\in[0, 2\pi]$. The parameter $\alpha^2$ which codify the presence of the global monopole is, in this case, given by \cite{Ola} $\alpha^2=1-\kappa^2\eta^2_0/2$ which is smaller than unity. In this coordinate system the metric tensor is explicitly defined as shown 
below:
\begin{eqnarray}
\label{Metric}
g_{44}=1, \  g_{55}=\alpha^2r^2, \ g_{66}&=&\alpha^2r^2\sin^2\theta_1, \ g_{77}=\alpha^2r^2\sin^2\theta_1\sin^2 \theta_2 \ . 
\end{eqnarray}
This bulk pacetime presents a scalar curvature $R=6(1-\alpha^2)/\alpha^2r^2$.

The scalar Green function associated with a massless field in this geometry must obey the non-homogeneous second order differential equation
\begin{equation}
\left(\Box-\xi R\right)G_E(x,x')=-\delta^n(x,x')=-\frac{\delta^{(p+4)}(x-x')}{\sqrt{g}} \ ,
\end{equation}
where we have admitted a non-minimal coupling between the field with the geometry. The d'Alembertian operator in the geometry defined by (\ref{Metric}) reads:
\begin{eqnarray}
	\Box=\partial_\mu\partial^\mu+\frac{\partial^2}{\partial r^2}+\frac3r\frac\partial{\partial r} -\frac{{\vec{L}}^2_{(3)}}{\alpha^2r^2} \ ,
\end{eqnarray}
where ${\vec{L}}^2_{(3)}$ is the angular operator defined in terms of the angular variables.

As in Section \ref{Scalar}, the scalar Green function can be obtained by calculating the heat kernel function, $K(x',x;s)$. The complete set of normalized eigen-function of the operator $\Box-\xi R$ is given by:
\begin{equation}
\Phi_\sigma(x)=\frac{{\sqrt{q}}}{\alpha^{3/2} (2\pi)^{2}}e^{-ikx}\frac{J_{\nu_l}(qr)}rY_{l,m_1,m_2}(\theta_1, \theta_2, \phi) \ ,
\end{equation}
with $\sigma^2=k^2+q^2$. In the function above, $\nu_l=\alpha^{-1}\sqrt{(l+1)^2+6(1-\alpha^2)(\xi-1/6)}$ and $Y_{l,m_1,m_2}(\theta_1, \theta_2, \phi)$ represents the hyperspherical harmonics of degree $l$ \cite{Y}, eigen-functions of the angular momentum operator ${\vec{L}}^2_{(3)}$ with eigen-values $l(l+2)$.

So according to (\ref{Heat}) and (\ref{Heat-1}) the scalar Green function is:
\begin{eqnarray}
\label{Green1}
G(x',x)=\frac{{\sqrt{2\pi}}}{2^4\pi^{5}\alpha^3}\frac{i}{(r'r)^3}\frac1{\sinh^{3/2}u}\sum_{l=0}^\infty(l+1)Q_{\nu_l-1}^{3/2}(\cosh u)C_l^1(\cos\gamma) \ , 	
\end{eqnarray}
being $C_l^n(x)$ the Gegenbauer polynomial of degree $l$ and order $n$.

Because we want to evaluate the Green function in the coincidence limit, we may take $\gamma=0$. Expressing the associated Legendre function in a polynomial form the Green function reads:
\begin{eqnarray}
\label{G1}	G(x',x)=\frac1{2^4\pi^4\alpha^3}\frac1{(r'r)^3}\frac1{\sinh^2u}\sum_{l=0}^\infty(l+1)^2\left[(\nu_l+1)e^{-\nu_lu}+\frac{e^{-(\nu_l+1)u}}{\sinh u}\right] \ .
\end{eqnarray}
The summation inside the bracket can be written as $S(u)\coth u-S'(u)$, where $S(u)$ is given by
\begin{eqnarray}
	S(u)=\sum_{l=0}^\infty(l+1)^2e^{-\nu_lu} \ ,
\end{eqnarray}
and the prime denotes the differentiation with respect to $u$.

Again it is possible to obtain an approximated expression to $\nu_l$ in a series powers of the parameter $\eta^2$:
\begin{eqnarray}
	\nu_l\approx(l+1)(1+\eta^2/2)+\frac{3(\xi-1/6)}{l+1}\eta^2+O(\eta^4) \ .
\end{eqnarray}
Admitting this situation we have:
\begin{eqnarray}
	S(u)=\frac{\cosh(u/2)}{4\sinh^3(u/2)}\left[1-\frac{3u\eta^2}{2\sinh u}\left(1+4\xi\sinh^2(u/2)\right)\right] +O(\eta^4) \ .
\end{eqnarray}
The complete expression to (\ref{G1}) is a long one and we shall not write it down; however it is possible to see that it presents the general form below:
\begin{eqnarray}	G(x',x)=\frac{1+3\eta^2/2}{2^7\pi^4}\frac1{(r'r)^3}\frac1{\sinh^6(u/2)}+\frac{1}{2^4\pi^4}\frac1{(r'r)^3}\frac1{\sinh^2 u}\left[{\cal{U}}_1(u)+\xi{\cal{U}}_2(u)\right]\eta^2 \ .
\end{eqnarray}

In \cite{Chr} is explicitly given the formal procedure to calculate the Hadamard function in a spacetime of even dimensions. We also shall not reproduce it here, we only write down the singular behavior of the Hadamard function in the eight dimensional spacetime:
\begin{eqnarray}
\label{H8}
	G_H(x',x)=\frac{\Delta^{1/2}(x',x)}{16\pi^4}\left[\frac{a_0(x',x)}{\sigma^3(x',x)}+\frac{a_1(x',x)}{4\sigma^2(x',x)}+ \frac{a_2(x',x)}{8\sigma(x',x)}-\frac{a_3(x',x)}{16}\ln\left(\frac{\mu^2\sigma(x',x)}2\right)\right] \ .
\end{eqnarray}

Because we are interested to calculate the renormalized vacuum expectation value of the square of the field operator up to the first order in the parameter $\eta^2$, we shall need the Hadamard function above up to the same order in the parameter $\eta^2$. In this case, the scalar curvature in this eight dimensional spacetime is $R=\frac{6\eta^2}{r^2}$. The functions that appear in (\ref{H8}), up to the first order in $\eta^2$, are
\begin{eqnarray}
	\Delta(x',x)&\approx& 1 \ , \nonumber\\
	 a_1(x',x)&\approx&\frac{6\eta^2}{r^2}\left(\frac16-\xi\right) \ .
\end{eqnarray}
Because, in this spacetime, $\Box R=0$, the coefficients $a_2(x',x)$ and $a_3(x',x)$ vanish, and the Hadamard function reads:
\begin{eqnarray}
	G_H(x',x)=\frac1{16\pi^4}\left[\frac1{\sigma^3(x',x)}+\frac{3\eta^2}{2r^2}\left(\frac16-\xi\right)\frac1{\sigma^2(x',x)}\right] \ .	
\end{eqnarray}

Now we are in position to calculate the vacuum average mentioned before. By using (\ref{Phi2}) we have:
\begin{eqnarray}
	\langle\Phi^2(x)\rangle_{Ren.}=\frac1{80\pi^4}\frac1{r^6}\left(\xi-\frac3{14}\right)\eta^2+O(\eta^4) \ .
\end{eqnarray}

It is interesting to notice that, although the dimension of the spacetime be even, there is no ambiguity in the vacuum average above up to the first order of $\eta^2$; moreover we can see that for the conformal coupling in eight dimensions, $\xi=\frac3{14}$, the vacuum expectation value above vanishes.

\end{document}